\documentclass[11pt]{article}
\usepackage[utf8]{inputenc}
\usepackage[T1]{fontenc}
\usepackage{amsmath}
\usepackage{amsfonts}
\usepackage{amssymb}
\usepackage{graphicx}
\usepackage{cite}
\usepackage[hyphens]{url}
\usepackage{hyperref}
\usepackage{algorithm}
\usepackage{algorithmic}
\usepackage{subcaption}
\usepackage{booktabs}
\usepackage{template}
\usepackage{balance}
\usepackage{xcolor}
\usepackage{tabularx}
\usepackage{listings}
\usepackage[section]{placeins}
\usepackage{stfloats}
\usepackage{pgfplots}
\usepackage{pgfplotstable}
\usepackage{enumitem}
\pgfplotsset{compat=1.18}
\usepackage[final]{microtype}
\hypersetup{hidelinks,breaklinks=true}
\allowdisplaybreaks
\setlist{nosep}
\setlist[itemize]{leftmargin=*,labelsep=0.5em}

\lstdefinelanguage{JavaScript}{
  keywords={typeof, new, true, false, catch, function, return, null, catch, switch, var, if, in, while, do, else, case, break, const, let, class, export, import, async, await},
  keywordstyle=\color{blue}\bfseries,
  ndkeywords={class, export, boolean, throw, implements, import, this},
  ndkeywordstyle=\color{darkgray}\bfseries,
  identifierstyle=\color{black},
  sensitive=false,
  comment=[l]{//},
  morecomment=[s]{/*}{*/},
  commentstyle=\color{purple}\ttfamily,
  stringstyle=\color{red}\ttfamily,
  morestring=[b]',
  morestring=[b]"
}

\renewcommand{\headeright}{Preprint; November 2025}
\renewcommand{\undertitle}{Preprint Version}
\renewcommand{\shorttitle}{AWARE: PriorityFresh Caching}

    \title{AWARE: Evaluating PriorityFresh Caching for Offline Emergency Warning Systems}

\author{
    	\textbf{Charles Melvin}\textsuperscript{1}, \textbf{N. Rich Nguyen}\textsuperscript{1} \\
    \small\textsuperscript{1}Department of Computer Science, University of Virginia, Charlottesville, VA 22903 \\
    \small\texttt{\{jgm6jy, nn4pj\}@virginia.edu}
}

\date{\small November 2025}

\microtypesetup{protrusion=true,expansion=true}

\begin{document}

\maketitle

\begin{abstract}
PriorityFresh is a semantic, actionability-first caching policy designed for offline emergency warning systems. Within AWARE’s simulation environment, PriorityFresh optimizes which alerts to retain and surface under constrained connectivity. Experiments indicate improved actionability-first performance without harming efficiency. A separate Priority Forecasting (PF) model is used only to synthesize realistic alert sequences for controlled experiments and does not influence caching or push decisions.
\end{abstract}


\section{Introduction}
Disaster-time information systems face a simple but consequential question: \emph{How can emergency systems most effectively deliver the most critical alerts to people?} In AWARE, which encompasses multiple components, “most pressing” refers to alerts that are both high-impact (severity, urgency) and still relevant (freshness). PriorityFresh is evaluated in isolation: it ranks content semantically and maintains top-priority items until they decay below relevance or expire. A separate Priority Forecasting (PF) model is used only upstream to help generate realistic alert scenarios for the simulator; PF and PriorityFresh are operationally independent—the cache policy does not consume PF outputs, and PF does not alter cache or push decisions.

\section{Contributions}
Key contributions (within a simulation-only evaluation):
\begin{itemize}
    \item \textbf{AWARE simulation environment.} A browser-hosted, offline-first simulator with reproducible, seeded environments and persistent run history for analysis—used to evaluate components of the broader AWARE system.
    \item \textbf{PriorityFresh policy.} A semantic, actionability-first eviction score is formalized.
    \item \textbf{Human-centered metrics.} Metrics that align with protective action (actionability-first, timeliness consistency, redundancy) are defined and reported, alongside standard efficiency (hit, delivery, freshness).
    \item \textbf{Empirical findings and operator guidance.} Across cache-size, network-reliability, joint sweeps, and extreme scenarios, PriorityFresh consistently improves actionability without harming efficiency; delivery is dominated by network physics. A winner-matrix view and a context-aware recommendation table summarize policy selection by device capability and network conditions, including cases where PAFTinyLFU is preferred (mission-critical first-surface regularity) and where TTLOnly is preferred (strict freshness SLAs), to avoid policy bias.
    \item \textbf{Stakeholder-facing implications (preview).} The evaluation later details role-specific guidance. In brief: (i) alert originators benefit from standardized CAP severity/urgency coding, thread identifiers, and tighter polygons with concise directives; (ii) telecom/platform providers can expose delivery telemetry, support device-assisted geofencing, and offer rate-limit/dedup primitives; (iii) app operators can match policy to objective (PriorityFresh under constrained caches or unreliable links; PAFTinyLFU when first-surface timing regularity is paramount; TTLOnly for strict freshness SLAs) and configure weights $(w_S,w_U,w_F)$ and push guardrails $R,D,\theta$ with fail-open treatment for Immediate/Extreme; (iv) local governments and community organizations can localize templates, validate relevance thresholds, and maintain fresh shelter data; (v) end users benefit from clear, low-noise interfaces that surface the most actionable item first and retain guidance for auditability.
\end{itemize}

\section{Related work}

\subsection{Emergency alert systems}
Modern public warning in the U.S. centers on FEMA's Integrated Public Alert and Warning System (IPAWS), which brokers alerts from authorized originators to distribution channels including Wireless Emergency Alerts (WEA), the broadcast Emergency Alert System (EAS), and NOAA Weather Radio \cite{nasem-2018-alerts}. IPAWS exchanges alert content using the Common Alerting Protocol (CAP) v1.2, enabling structured, multi-channel delivery and machine-readable metadata \cite{oasis-cap-1.2}. Since 2018, successive FCC rulemakings have tightened WEA performance requirements: device-based geo\-targeting must match the target area with no more than approximately 0.1~mile overshoot beyond the polygon, and providers must meet transmission speed and logging requirements \cite{fcc-2018-geo,fcc-wea-2023-doc}. Empirically, independent assessments demonstrate strong but nonuniform reach and latency. RAND/HSOAC's national survey of the Oct.\ 4, 2023 nationwide test characterizes receipt, opt-in, and demographic patterns \cite{rand-wea-2023-test}, while McBride et~al. measured median delivery latencies on the order of seconds and largely accurate geofences in ShakeAlert-supported trials \cite{mcbride-2023-wea-latency}. These results motivate client-side resilience---e.g., prefetching, caching, and offline fallback---to hedge against residual gaps from congestion, RF shadowing, or localized outages.

\subsection{Offline-first applications}
Offline-first design patterns let safety-critical apps remain usable under partial connectivity. Local-first software research demonstrates how conflict-free replicated data types and opportunistic synchronization preserve functionality when connectivity is intermittent, providing a foundation for resilient client caches and collaborative state \cite{kleppmann-2019-localfirst}. In emergency contexts, these findings motivate prefetch of hazard layers, durable storage of prior alerts for auditability, and continuity of core functions during network degradation.

\subsection{Caching strategies for emergency systems}
Network-side caching can reduce tail latency and backhaul dependence during incident-driven traffic spikes. Surveys of mobile edge computing (MEC) detail cache placement and cooperation at the edge to improve hit rates and responsiveness for time-sensitive content \cite{mec-caching-survey-2023}. In Information-/Named-Data Networking (ICN/NDN), systematic surveys cover in-network caching strategies tuned to popularity, mobility, and energy constraints, including IoT deployments that resemble disrupted, ad hoc post-disaster networks \cite{icn-iot-caching-survey-2023}. Recent techniques explore learning-based placement (e.g., bandit/RL hybrids) and hybrid popularity/centrality policies to sustain performance under mobility and intermittent links \cite{cache-mab-2023,electronics-2024-koide-icanet}. Beyond hit ratio, "freshness" is a first-order requirement for real-time guidance; system-level treatments frame adaptive TTLs, invalidation, and priority refresh as design knobs that trade staleness against load and reach \cite{hotnets-2024-freshness}. AWARE adopts these lessons with priority-aware eviction and prefetch keyed to hazard severity, temporal sensitivity, and proximity.

\subsection{How PriorityFresh relates to prior caching}
Classical eviction and admission policies primarily optimize for recency/frequency and sometimes object cost/size: LRU/LFU (and spectra like LRFU) and adaptive policies like ARC improve reuse without explicit content semantics; cost-aware policies such as GreedyDual-Size (GDS) incorporate retrieval cost and size into a single eviction score \cite{cao-irani-1997-usits,megiddo-modha-2003-arc}. More recent work separates admission from eviction, e.g., TinyLFU filters low-utility items before they pollute the cache \cite{einziger-2017-wtinylfu}. Parallel threads study freshness and timeliness trade-offs, treating staleness as a controllable resource (e.g., adaptive TTLs, refresh prioritization) for real-time content \cite{hotnets-2024-freshness}.

In disrupted or delay-tolerant settings, prioritization has been explored at the message level: DTN protocols like RAPID maximize delivery utility under bandwidth/contact constraints using urgency/deadline-like objectives, with drop/replication policies driven by per-message utility rather than access history \cite{balasubramanian-2007-rapid}. ICN/NDN literature also includes popularity- or class-based priorities, but typically lacks end-user hazard semantics.

PriorityFresh differs in three ways:
\begin{itemize}
	\item Semantic, actionability-first scoring: eviction priority is a weighted, exponentially decayed function of hazard severity, temporal urgency, and freshness—variables tied to protective action rather than access counts or byte cost. Spatial proximity and deduplication are handled by geofencing and push admission/dedup rules, not the core score.
    \item Fixed, transparent weights (no learned boost): unlike learning-based cache policies, PriorityFresh does not consume ML signals; weights are operator-tunable and held fixed for all experiments.
    \item Push coordination under device constraints (optional): when enabled, push admission is treated as a rule-based guard with rate limit and dedup; in reported experiments pushes were disabled.
\end{itemize}
Together, these aspects target the emergency-warning objective directly—maximizing timely presentation of the single most actionable item for a user in place—rather than generic byte-hit or request-hit metrics.

\subsection{Location-based emergency services}
Recent FCC rulemakings now expect polygonal targeting with device-assisted geofencing in WEA, minimizing over-alerting while maintaining high coverage inside the alert area \cite{fcc-2018-geo,fcc-wea-2023-doc}. Empirical geofencing studies quantify precision and recall as a function of radius, environment, operating system, and event type, informing practical buffer selection and distance-weighted ranking in location-aware filtering \cite{shevchenko-2023-geofencing}. These threads motivate a combination of device-level geofencing with cache priorities that elevate locally relevant, high-impact content during flood scenarios.

\subsection{Human and cognitive factors in emergency communication}
Decades of risk-communication research show that the efficacy of alerts depends on clarity, timing, trust, and cognitive load---not delivery alone. Classic work and subsequent syntheses emphasize concise, directive messages and pathways from warning receipt to protective action (e.g., the Protective Action Decision Model) \cite{mileti-1990-ornl6609,lindell-2012-padm,nasem-2018-alerts}. During disasters, social support and perceived control correlate with better psychological outcomes; resilient communication systems can buffer stress by maintaining a sense of connection even asynchronously \cite{cohen-1985-socialsupport,norris-2008-resilience}. For interface design, trust and progressive disclosure support comprehension and compliance under stress \cite{paton-2008-warningresponse}. AWARE aligns with these principles by prioritizing terse, high-impact messages; surfacing nearby shelters and actionable guidance first; and retaining background data for audit and continuity when connectivity is degraded.

\section{Limitations of current weather alert delivery in practice}
\label{sec:limitations-weather}
Commercial weather services and consumer weather apps play an important role in disseminating severe weather warnings, but their delivery choices can be sub-optimal for end users in high-stress scenarios. The issues below, observed across public systems and commercial pipelines, motivate AWARE's actionability-first, precision-oriented design.

\subsection{Mass-Coverage Bias and Over-Alerting}
To maximize reach and manage liability, alerting systems and downstream distributors have historically opted for wider-than-necessary targeting (e.g., county-wide or buffered polygons), which increases the likelihood that users outside the immediate hazard area receive alerts. The emergency-alerting community has moved toward more precise, device-assisted geotargeting for this reason; U.S. rules for Wireless Emergency Alerts (WEA) explicitly require matching the target area to within approximately 0.1\,mile of overshoot, and codify performance, logging, and speed requirements \cite{fcc-2018-geo,fcc-wea-2023-doc}. The National Academies also note that over-alerting and excessive false positives erode trust and compliance over time \cite{nasem-2018-alerts}. Empirical work in severe weather corroborates this dynamic: false alarms are associated with diminished protective response and trust in future warnings \cite{simmons-2009-falsealarms,ripberger-2015-falsealarm}, and reviews of the tornado warning process highlight persistent challenges in reducing false-alarm-driven fatigue while maintaining detection performance \cite{brotzge-2013-tornado}.

\subsection{Duplicate Updates and Staleness}
Hazard evolutions are often communicated via a sequence of CAP updates (extended expirations, upgraded/downgraded severities, corrected polygons). When distribution pipelines surface each update as a new push without threading or deduplication, users experience alert storms and stale repeats, increasing cognitive load \cite{nasem-2018-alerts}. Greedy “first-in, first-out” queuing of updates can prioritize volume over value, surfacing low-utility corrections ahead of critical escalations. Some mobile platforms may also coalesce notifications for a bundle when delivery is deferred, which can change end-user perception of update frequency and ordering \cite{apple-apns-remote}.

\subsection{Geofencing Precision Varies by Platform and Context}
Even with polygonal targets, app-side geofencing precision depends on device OS, location providers, and environmental factors (e.g., urban canyons), leading to both over- and under-inclusion at the boundary \cite{shevchenko-2023-geofencing}. Field measurements of WEA delivery during ShakeAlert trials show generally low latencies but also platform-dependent variance \cite{mcbride-2023-wea-latency}, underscoring that end-to-end behavior is heterogeneous across devices and networks.

\subsection{Latency, Congestion, and Last-Mile Variability}
During incident-driven surges, back-end queuing, mobile-network congestion, and cloud push-notification throttling can introduce variable delays. Regulatory targets focus on WEA transmission speed \cite{fcc-wea-2023-doc}, while app ecosystems rely on best-effort push services that are subject to device power management and platform policies. On Android, devices in Doze may defer normal-priority messages until maintenance windows, while high-priority messages attempt immediate delivery but can still be deprioritized based on prior behavior \cite{android-doze-standby,fcm-android-priority}. FCM also documents conditions under which messages might not be delivered (e.g., long offline intervals, excessive pending queue) \cite{fcm-receive}. On Apple platforms, APNs may coalesce notifications for a bundle when immediate delivery isn’t possible \cite{apple-apns-remote}. These realities can cause late arrival or misordered updates relative to the true event timeline.

\subsection{User Trust and Cognitive Load}
Over-alerting, duplicates, and stale content increase cognitive load and diminish trust \cite{mileti-1990-ornl6609,nasem-2018-alerts,paton-2008-warningresponse}. Studies of severe-weather warnings specifically document how high false-alarm rates can suppress compliance with subsequent warnings \cite{simmons-2009-falsealarms,ripberger-2015-falsealarm}. In practice, people benefit from concise, directive messages that surface the most actionable items first and suppress redundant noise. AWARE's PriorityFresh and push-dedup rules directly address these failure modes by prioritizing urgency/severity with freshness, threading updates, and suppressing short-window repeats.

\paragraph{Implication.} Precision targeting, duplicate-aware threading, and actionability-first ranking are not merely engineering niceties; they are necessary conditions for effective public response. The evaluation therefore emphasizes metrics beyond delivery (e.g., actionability-first and timeliness consistency) to align system behavior with human factors.

\section{Smart caching algorithm}
AWARE includes the cache policies implemented in \texttt{src/sim/policies}.

\subsection{Policies available}
\begin{itemize}
    \item \textbf{LRU} evicts the least recently accessed alert after purging expired entries.
    \item \textbf{TTLOnly} keeps alerts until they expire, without considering recency.
    \item \textbf{PriorityFresh} scores alerts by semantic priority and freshness before eviction.
    \item \textbf{PAFTinyLFU} combines recency with a TinyLFU frequency sketch to admit only likely high-utility alerts.
\end{itemize}

\subsection{PriorityFresh scoring}
The PriorityFresh policy maintains an in-memory map with capacity $C$ and assigns each alert $a$ the score
$$
\mathrm{base}(a,t) = w_S \, s(a) + w_U \, u(a) + w_F \, f(a,t),
$$
where $s(a)$ and $u(a)$ are ordinal encodings (in code: $s(\text{Minor}){=}1,\ s(\text{Moderate}){=}2,\ s(\text{Severe}){=}3,\ s(\text{Extreme}){=}4,\ \\s(\text{Unknown}){\approx}2$ and $u(\text{Past}){=}0.5,\ u(\text{Future}){=}1.5,\ u(\text{Expected}){=}2,\ u(\text{Immediate}){=}3,\ u(\text{Unknown}){\approx}1.5$), and $f(a,t) = e^{-\lambda (t - a_{\mathrm{issued}})}$ applies exponential decay with $\lambda = 1/600$. This choice yields a priority half-life of $\ln 2 / \lambda \approx 416$\,s. The default \emph{ordering} emphasizes urgency and freshness over severity (\,$U\!\approx\!F\!>\!S$\,). In code, defaults are $w_S{=}2$, $w_U{=}3$, $w_F{=}4$. For the experiments reported in this paper we used $w_S{=}4$, $w_U{=}5$, $w_F{=}5$, which preserves the same ordering; absolute scaling of all three weights does not change eviction order, whereas the \emph{relative} magnitudes do. This choice aligns with protective-action guidance: timeliness and recency should keep an immediate, fresh update ahead of an older, nominally severe item. Operators can adjust the weights per run to fit context; a full sensitivity sweep is left for future work. When the cache is full, the alert with the lowest total score is evicted before inserting the newcomer. Expired alerts are purged on every read and write.
For query-side retrieval, the simulator samples cached items with an urgency-first bias. All results reflect fixed weights without ML inputs.

\subsection{Push notification optimization}
\label{sec:push-optimization}
When enabled, push delivery is modeled as a rule-based decision at alert arrival time under two operational constraints: a rate limit $R$ pushes per minute and a deduplication window $D$ seconds that suppresses rapid repeats from the same thread. Given alert $a$ at time $t$, a push occurs only if all of the following hold:
\begin{enumerate}
    \item within the rate limit (fewer than $R$ pushes in the last 60 seconds),
    \item not a recent duplicate (no push for the same \texttt{threadKey} within $D$ seconds), and
    \item $\mathrm{base}(a,t) \ge \theta$ or $a$ is high-impact (Extreme or Immediate).
\end{enumerate}
Here $\theta$ is an operator-tunable threshold. In the experiments reported, push delivery was disabled ($R {=} 0$).

\begin{algorithm}[t]
\caption{PriorityFresh cache update and optional push decision.}
\label{alg:priorityfresh}
\begin{algorithmic}[1]
\STATE purgeExpired($t$)
\STATE base $\leftarrow w_S s(a)+ w_U u(a)+ w_F f(a,t)$
\IF{cache is full} \STATE evict min-score entry \ENDIF
\STATE insert $a$ with score base
\IF{within rate limit AND not duplicate AND $(\text{base} \ge \theta$ OR high-impact$)$}
    \STATE send push; record timestamp and thread
\ELSE
    \STATE suppress push; record suppression reason
\ENDIF
\end{algorithmic}
\end{algorithm}

\section{System architecture}
AWARE is a larger system concept; a browser-hosted single-page simulation environment enables testing and analysis of components (including PriorityFresh) before any deployment. It is composed of the following layers:

\begin{enumerate}
    \item \textbf{Interaction layer (React/TypeScript).} \texttt{App.tsx} coordinates simulation controls, results, run history, active alerts, shelter listings, and the environment view. A new \texttt{SimulationProvider} wraps the component tree so each tab consumes a shared snapshot of the latest run.
    \item \textbf{Simulation core.} \texttt{runSimulation} in \texttt{src/sim/run.ts} derives a seeded environment, synthesizes alert streams for controlled experiments, executes the selected cache policy, and emits metrics plus per-region delivery statistics. Scenario specifications in \texttt{src/sim/scenarios/*.ts} describe alert rates, outages, and target first-delivery service level agreements, while all randomness flows through a deterministic Mulberry32 generator.
    \item \textbf{Persistence layer (Dexie/IndexedDB).} \texttt{src/db.ts} defines the \texttt{AwareDB} schema with object stores for reports, shelters, runs, and key--value metadata. Helpers such as \texttt{putReports}, \texttt{putShelters}, and \texttt{logRun} persist the full run result so it can be replayed without re-executing the simulator.
    \item \textbf{Context and data providers.} The environment generator in \texttt{src/sim/geo/generate.ts} turns the seeded RNG into polygonal regions with local reliability multipliers rendered by \texttt{EnvironmentView.tsx}. Static shelter data and service wrappers under \texttt{src/api} expose Weather and Mapbox readiness, and the Services tab now confirms credentials by rendering a Mapbox static preview.
\end{enumerate}

\noindent\textbf{Execution flow.} When the operator triggers "Run Simulation," the UI locks controls, invokes \texttt{runSimulation(options)}, and receives a payload that now bundles alerts, metrics, the seeded environment, and per-region stats. The snapshot is persisted to Dexie and injected into the shared \texttt{SimulationContext}, so history, active alerts, Results, and the Environment tabs all reference the same run without re-executing the core.

\subsection{Data model}
The persisted data structures mirror the TypeScript types in \texttt{src/db.ts}.

\subsubsection{Emergency reports}
Each report approximates a CAP message subset:
\begin{itemize}
    \item \texttt{id}: unique identifier generated by the simulator.
    \item \texttt{eventType}: coarse category such as \texttt{Flood} or \texttt{Shelter}.
    \item \texttt{severity} $\in \{\texttt{Minor}, \texttt{Moderate}, \texttt{Severe},\\ \texttt{Extreme}, \texttt{Unknown}\}$ and \texttt{urgency} $\in \{\texttt{Immediate}, \texttt{Expected},\\ \texttt{Future}, \texttt{Past}, \texttt{Unknown}\}$.
    \item \texttt{issuedAt} / \texttt{expiresAt}: Unix seconds used for ``active now'' filtering.
    \item Optional metadata: \texttt{headline}, \texttt{instruction}, \texttt{sizeBytes}, \texttt{geokey}, and \texttt{polygon}.
\end{itemize}

\subsubsection{Shelter information}
Shelter rows describe nearby options surfaced in the UI:
\begin{itemize}
    \item \texttt{id}, \texttt{name}, and optional \texttt{address}.
    \item \texttt{coordinates}: longitude/latitude pairs for map integration.
    \item \texttt{capacity} and \texttt{status} $\in \{\texttt{open}, \texttt{full}, \texttt{closed}\}$.
    \item \texttt{updatedAt} timestamps and optional \texttt{geokey}.
\end{itemize}

\subsubsection{Run metadata}
Runs capture provenance for reproducibility:
\begin{itemize}
    \item \texttt{id}, \texttt{scenario}, \texttt{policy}, \texttt{seed}, and \texttt{timestamp}.
    \item Serialized \texttt{metrics}, \texttt{samplesCount}, and optional \texttt{experimentName}, \texttt{notes}, \texttt{fullResults}.
\end{itemize}

\subsection{Plain-language interpretation}
To make the simulator legible for non-technical stakeholders, core terms map to everyday meaning:
\begin{itemize}
    \item \textbf{Alert reports} are the notifications a resident sees (e.g., "Flash Flood Warning") with headline and instructions kept for offline reference.
    \item \textbf{Polygons and geokeys} approximate neighborhoods: polygons sketch target areas; geokeys are coarse grid labels for grouping nearby locations without full maps.
    \item \textbf{Cache policies} decide which alerts a handset keeps locally so guidance remains available when connectivity drops.
    \item \textbf{Freshness/TTL} is the "expires after" time that marks when guidance goes out of date.
    \item \textbf{Seeds, modes, and replicates} are the recipe for reproducibility: seeds replay the same timeline, seed modes add controlled variation, and replicates repeat runs to check consistency.
    \item \textbf{Metrics} (hit rate, redundancy, timeliness consistency, actionability-first) proxy common questions: Did people get it? Was it noisy? Did critical updates arrive on time?
    \item \textbf{Dexie/IndexedDB} is the browser's local database holding alerts, shelters, and audit logs so the interface works offline.
    \item \textbf{Environment providers} (weather, Mapbox) are future data feeds; the UI shows API-key readiness for live integration.
\end{itemize}

\subsection{Database schema}
Dexie manages the IndexedDB stores via the following definition:

\begin{lstlisting}[language=JavaScript, caption={AWARE database definition in `src/db.ts`}]
export class AwareDB extends Dexie {
  reports!: Table<Report, string>;
  shelters!: Table<Shelter, string>;
  runs!: Table<RunMeta, string>;
  kvs!: Table<KV, string>;

  constructor() {
    super('awareDB');
    this.version(1).stores({
      reports: 'id, issuedAt, expiresAt, severity, urgency, geokey, eventType',
      shelters: 'id, geokey, status, updatedAt, name',
      runs: 'id, timestamp, scenario, policy, seed, experimentName',
      kvs: 'key'
    });
  }
}
\end{lstlisting}

\section{Metrics}
The following metrics are used consistently across experiments; thresholds and windows are fixed per batch for comparability.
\begin{itemize}
    \item \textbf{Delivery rate:} fraction of alerts successfully delivered to the device under simulated reliability and retry settings.
    \item \textbf{Cache hit rate:} fraction of retrievals served from the local cache.
    \item \textbf{Average freshness:} mean of the normalized freshness $f(a,t)$ at evaluation time, using the same decay as in PriorityFresh.
    \item \textbf{Stale access rate:} fraction of retrievals whose item has expired (i.e., $t>\texttt{expiresAt}$).
    \item \textbf{Actionability-first ratio:} fraction of threads for which the first surfaced item satisfies the fixed actionability predicate used in all experiments: an alert is actionable iff (i) \texttt{urgency} $=\,$\texttt{Immediate} or (ii) \texttt{severity} $\in\{\,$\texttt{Severe}, \texttt{Extreme}\,$\}$. A thread is defined by \texttt{threadKey} (or \texttt{id} if absent), and the "first surfaced item" is the first successful retrieval from the cache for that thread during the run. This predicate does not depend on freshness/TTL or PF outputs and is held constant across policies for comparability.
    \item \textbf{Timeliness consistency:} fraction of threads whose first surface occurs within a fixed time window $\delta$ of the item’s issue time.
    \item \textbf{Redundancy (if reported):} rate of duplicate surfaces within a short window for the same thread.
\end{itemize}

\section{Experimental setup}
The experimental configuration and visual assets summarize the test runs. Brief interpretations accompany the figures for clarity; figures remain primary.

\subsection{Default parameters}
Unless otherwise specified, all single-seed comparisons use the following settings:

\begin{table}[ht]
\centering
\caption{Default run configuration}
\label{tab:default-config}
\begin{tabular}{@{}ll@{}}
    \toprule
Scenario & Urban \\
Cache Size & 128 \\
Alerts & 400 \\
Severity weight $w_S$ & 4 \\
Urgency weight $w_U$ & 5 \\
Freshness weight $w_F$ & 5 \\
Network reliability & 0.85 \\
Seed & FISHDINNER \\
Seed mode & fixed (single seed) \\
Replicates & 1 \\
Duration (sec) & 900 \\
Query rate (per minute) & 60 \\
\bottomrule
\end{tabular}
\end{table}

\noindent Delivery behavior for single-seed (non-randomized) runs:

\begin{table}[ht]
\centering
\caption{Delivery settings (single-seed runs only)}
\label{tab:delivery-settings}
\begin{tabular}{@{}ll@{}}
   \toprule
Retry interval (sec) & 1 \\
Max attempts & 10 \\
\bottomrule
\end{tabular}
\end{table}

\subsection{Policies compared}
All experiments compare four cache policies under identical seeds and controls unless noted: LRU, TTLOnly, PriorityFresh, and PAFTinyLFU.

\subsection{Batch designs}
Four batches were executed:
\begin{enumerate}
        \item \textbf{Baseline comparison (fixed seed and config).} All four policies evaluated under the default parameters in Table~\ref{tab:default-config}.
        \item \textbf{Cache-size sweep.} All four policies evaluated over multiple cache capacities with all other parameters fixed.
        \item \textbf{Network-reliability sweep.} All four policies evaluated across varying baseline reliabilities with all other parameters fixed.
        \item \textbf{Joint sweep (cache size \& network).} All four policies evaluated over a grid of cache capacities and reliabilities.
\end{enumerate}

\section{Results}
\label{sec:results}
Quantitative findings from the CSV outputs in \texttt{scripts/data} are summarized here. Unless noted, all runs use the default configuration in Table~\ref{tab:default-config}. Pushes were disabled (rate limit $R{=}0$), so delivery is governed solely by network reliability and cache behavior.

\subsection{Baseline (single seed)}
\begin{figure}[!t]
    \centering
    \includegraphics[width=\textwidth]{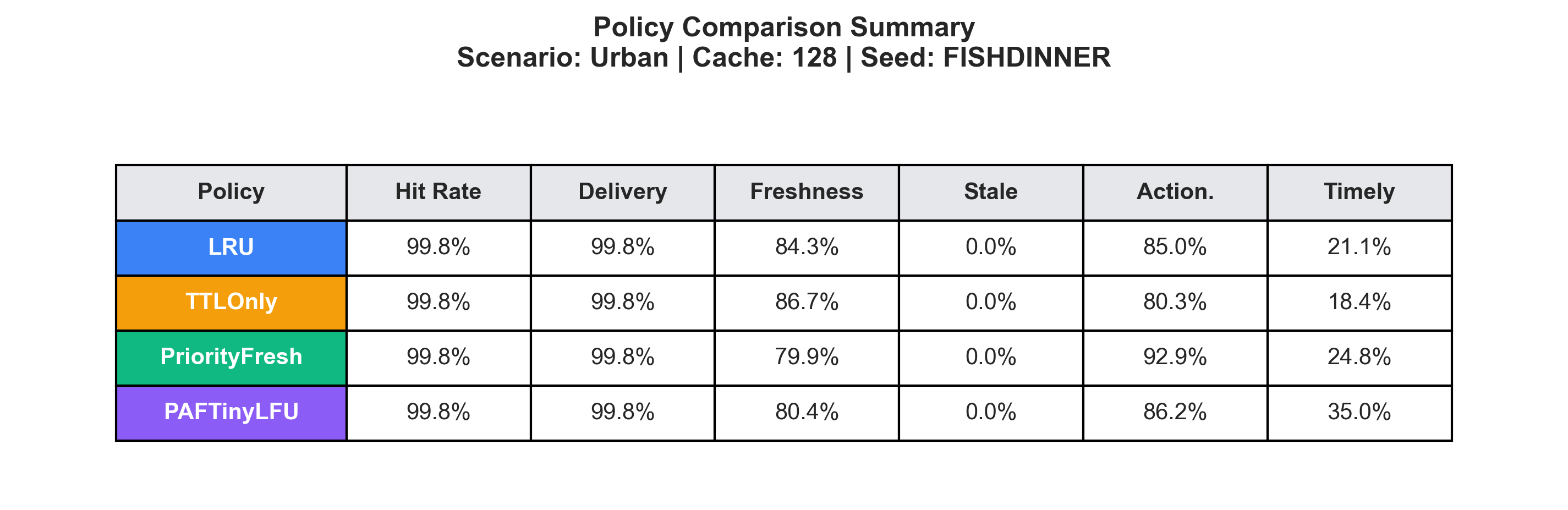}
    \caption{Summary table of baseline metrics across policies.}
    \label{fig:baseline-table}
\end{figure}

All four policies achieved nearly identical system efficiency under the baseline seed: cache hit rate $\approx 0.998$ and delivery rate $\approx 0.998$. Differences emerge in human-centered metrics. The summary table is shown as an image (Fig.~\ref{fig:baseline-table}) rather than a LaTeX table for consistency with the rest of the composite figures:

\noindent Observations (supported by Fig.~\ref{fig:baseline-table}):
\begin{itemize}
    \item \textbf{Actionability-first.} PriorityFresh leads on the fraction of threads where the first surfaced item is actionable, reflecting its semantic priority design (no ML boost used).
    \item \textbf{Freshness.} TTLOnly yields the freshest content (as expected from a TTL policy). PriorityFresh trades a small amount of freshness for actionability-first.
    \item \textbf{Timeliness consistency.} PAFTinyLFU stabilizes first-push timeliness the most; PriorityFresh improves over LRU/TTL but generally trails PAFTinyLFU on this metric.
\end{itemize}

\FloatBarrier

\subsubsection*{Detailed baseline analysis}
	\textbf{Cache efficiency (hit \& delivery).} All policies attain near-identical hit and delivery ($\approx$99.8\%), implying the cache is sufficiently provisioned under this workload and that qualitative differences arise from which items are prioritized and refreshed rather than from misses.

	\textbf{Freshness.} TTLOnly yields the highest average freshness ($\sim$86.7\%), as expected from enforcing TTL-driven retention/refresh. PriorityFresh trades some freshness ($\sim$79.9\%) to favor semantically critical content; LRU and PAFTinyLFU sit between these extremes.

	\textbf{Staleness.} All policies record 0\% stale, indicating expiry/refresh controls are well-tuned for this baseline.

	\textbf{Actionability-first.} PriorityFresh leads ($\sim$92.9\%), reflecting its design goal to surface high-impact alerts first. TTLOnly is lowest ($\sim$80.3\%), consistent with uniform freshness emphasis over semantic priority. LRU and PAFTinyLFU deliver mid-80s by balancing recency/frequency without explicit hazard semantics.

	\textbf{Timeliness consistency.} PAFTinyLFU leads (\,$\sim$35\%\,), suggesting that its frequency-aware admission produces more stable first-surface timing. PriorityFresh is above LRU/TTL (\,$\sim$24.8\%\,) but trails PAFTinyLFU; TTLOnly is lowest (\,$\sim$18.4\%\,), likely due to time-based resets that incur overhead without prioritization.

\subsubsection*{Policy tradeoffs in the baseline}
\begin{itemize}
    \item \textbf{LRU:} Simple, consistent baseline with solid overall efficiency; lacks prioritization and shows average timeliness behavior.
    \item \textbf{TTLOnly:} Best freshness and easy to tune; underperforms on actionability-first and timeliness consistency.
    \item \textbf{PriorityFresh:} Highest actionability-first aligned with human-centered objectives; accepts a modest freshness trade and moderate timing stability.
    \item \textbf{PAFTinyLFU:} Best timeliness consistency and strong balance overall; freshness slightly below TTLOnly but competitive.
\end{itemize}

\noindent\textbf{Bottom line.} Under this cache scale and workload, PriorityFresh achieves its objective of surfacing the most actionable content first without degrading system efficiency; PAFTinyLFU is preferred when consistent first-surface timing is paramount; TTLOnly is a freshness-first baseline.

\subsection{Cache-size scaling (device profiles)}
\begin{figure}[!t]
    \centering
    \includegraphics[width=\linewidth]{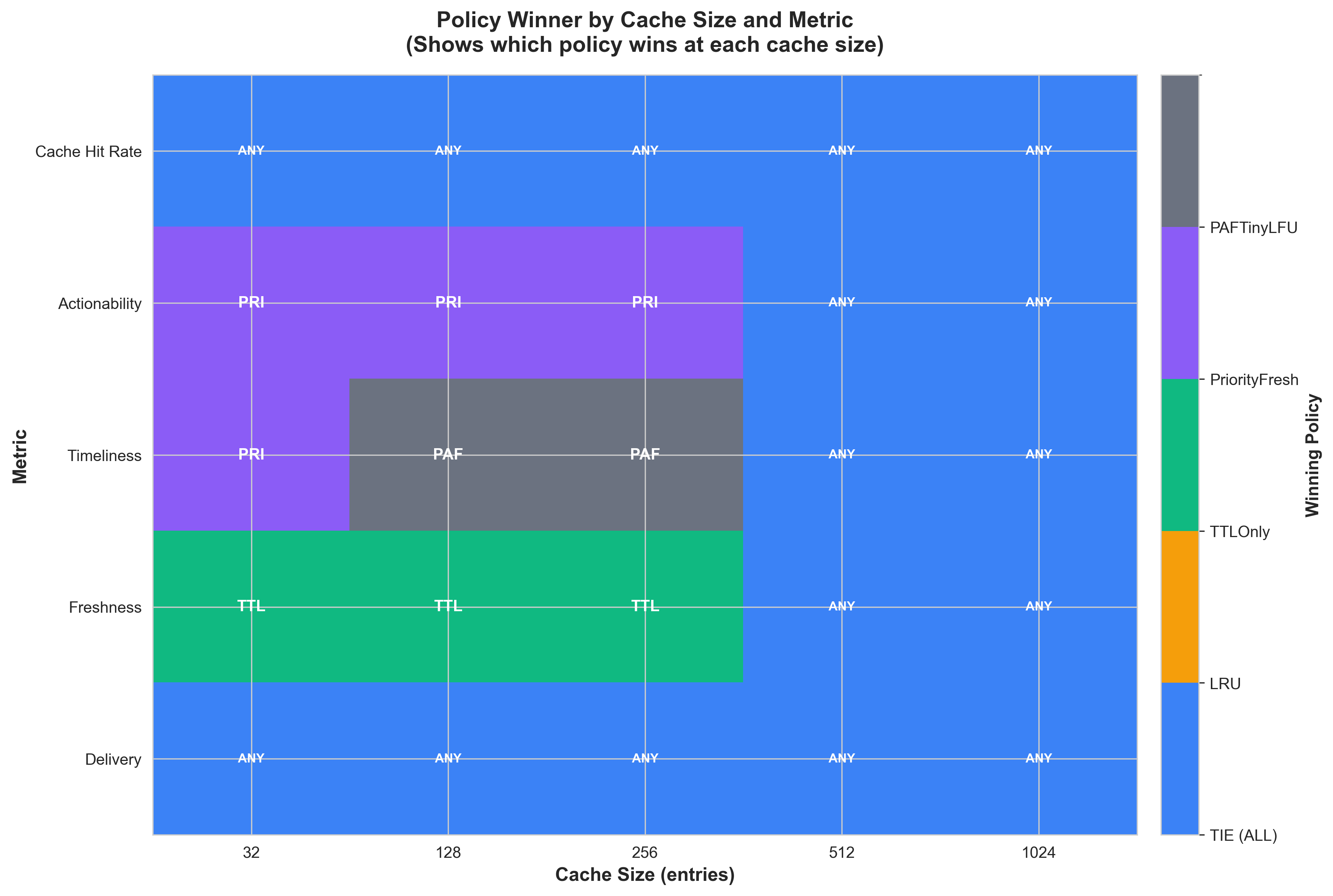}
    \caption{Cache-size sweep: winner heatmap across metrics/policies.}
    \label{fig:device-winner-heatmap}
\end{figure}

\begin{figure}[!t]
    \centering
    \includegraphics[width=\textwidth]{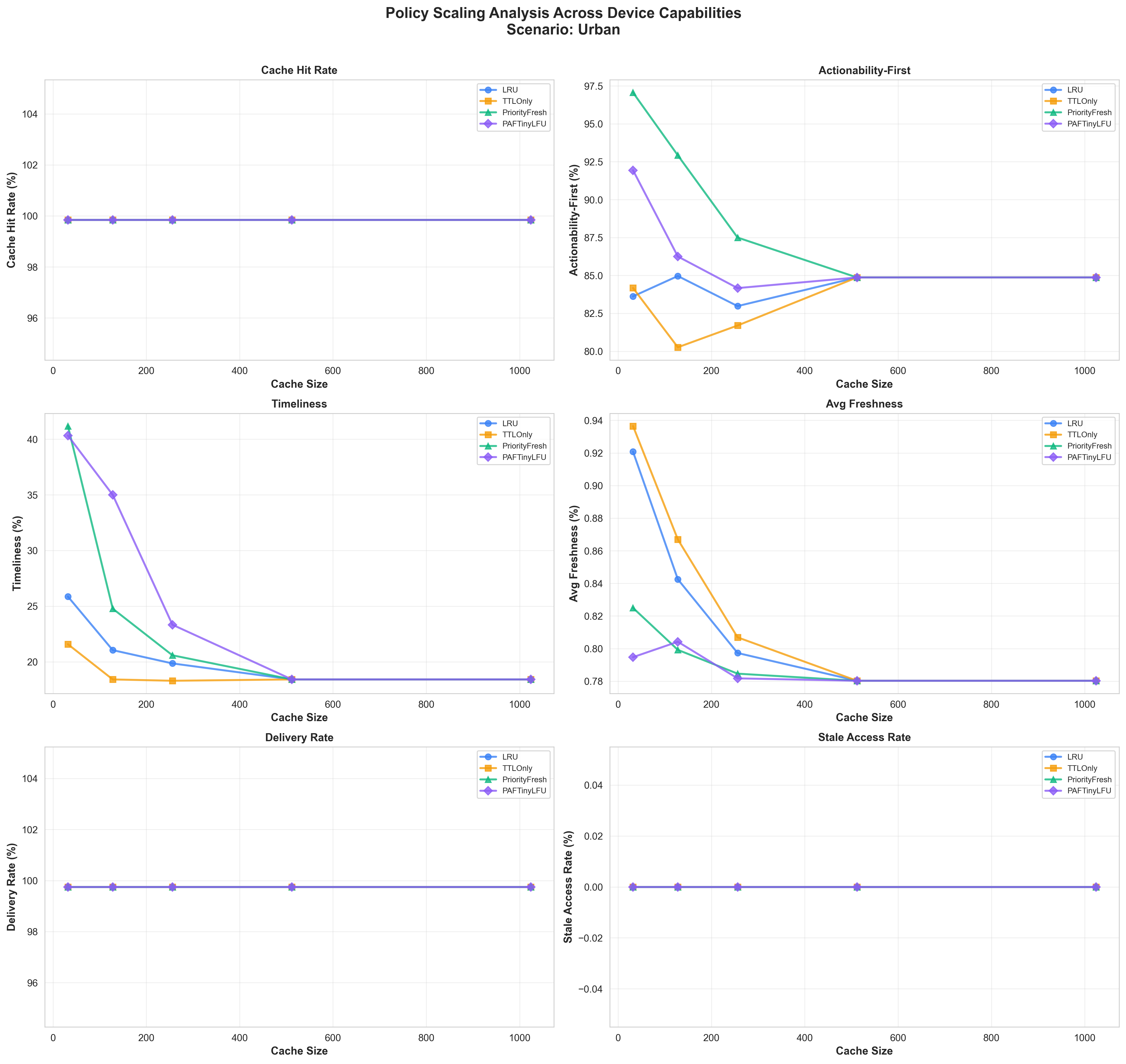}
    \caption{Cache-size sweep: all-metrics grid overview.}
    \label{fig:device-all-grid}
\end{figure}

From \texttt{device-comparison-*.csv} and Fig.~\ref{fig:device-winner-heatmap} and Fig.~\ref{fig:device-all-grid}:
\begin{itemize}
    \item \textbf{Small caches (e.g., 32 entries, Budget Phone).} PriorityFresh maintains the highest actionability-first ratio ($\approx$0.97) and the highest timeliness consistency in this tier, while TTLOnly/LRU exhibit higher average freshness.
    \item \textbf{Moderate caches (128--256).} PriorityFresh continues to lead on actionability-first; PAFTinyLFU often edges out on timeliness consistency; freshness ordering remains TTLOnly $>$ LRU $>$ PriorityFresh $\approx$ PAFTinyLFU.
    \item \textbf{Large caches (512+).} Policies converge across all reported metrics—capacity is sufficient to retain most alerts, making selection effects negligible.
\end{itemize}

\subsubsection*{Policy scaling across device capabilities (Fig.~\ref{fig:device-all-grid})}
	\textbf{Scenario.} Urban. The all-metrics grid summarizes how LRU, TTLOnly, PriorityFresh, and PAFTinyLFU scale with increasing cache size, capturing resource-dependent and invariant behaviors across six primary metrics; the winner heatmap (Fig.~\ref{fig:device-winner-heatmap}) provides a complementary, at-a-glance summary.

	\textbf{Cache hit rate.} Across 32\,$\to$\,1024 entries, hit rate stays effectively constant near $\sim$99.8\% for every policy. The working set fits even the smallest cache, so eviction details are negligible; above the footprint, hit rate saturates.

	\textbf{Actionability-first.} Divergence is largest at small capacities ($\le$256): PriorityFresh leads (\,$\sim$93--97\%\,), reflecting its bias toward mission-critical alerts under scarcity. LRU/TTLOnly start \,$\sim$82--85\%\, and converge toward parity as capacity grows. By 512+ entries, all flatten near \,$\sim$85\%\,, indicating early differences are resource-driven, not algorithmic at scale.

	\textbf{Timeliness.} Timeliness shows the steepest scaling gradient: PAFTinyLFU begins near \,$\sim$40\%\, (best short-latency consistency) but decays toward \,$\sim$20\%\, as caches grow. PriorityFresh maintains a small edge over LRU/TTLOnly at low capacities, then converges. Temporal advantages diminish once the system stops trading freshness vs. latency, reaching caching equilibrium at high capacity.

	\textbf{Average freshness.} TTLOnly is strongest at small caches due to enforced TTLs. All policies converge by 512--1024 entries to $\approx 0.80$--$0.82$ normalized freshness as turnover slows and policy differences become marginal.

	\textbf{Delivery rate.} Flat and near-perfect (\,$\sim$99.8\%\,) across sizes and policies, implying network conditions dominate delivery under Urban reliability; caching choices have little effect here.

	\textbf{Stale access rate.} All policies maintain 0\% across all cache sizes, confirming no propagation of invalid data and consistent refresh behavior during scaling.

	\textbf{Interpretation.} These trends demonstrate diminishing returns from larger caches: under constraint, policy logic matters—PriorityFresh improves actionability, PAFTinyLFU enhances timeliness, TTLOnly sustains freshness. As capacity increases, the system enters a saturation regime where objects persist and strategic replacement loses impact. Policy intelligence thus matters most under constraint; once capacity (and stability) dominate, context-aware differentiation collapses into parity, consistent with the \emph{ANY} patterns in the winner summaries.
\FloatBarrier

\subsection{Network reliability sweep}
\begin{figure}[!t]
    \centering
    \includegraphics[width=\linewidth]{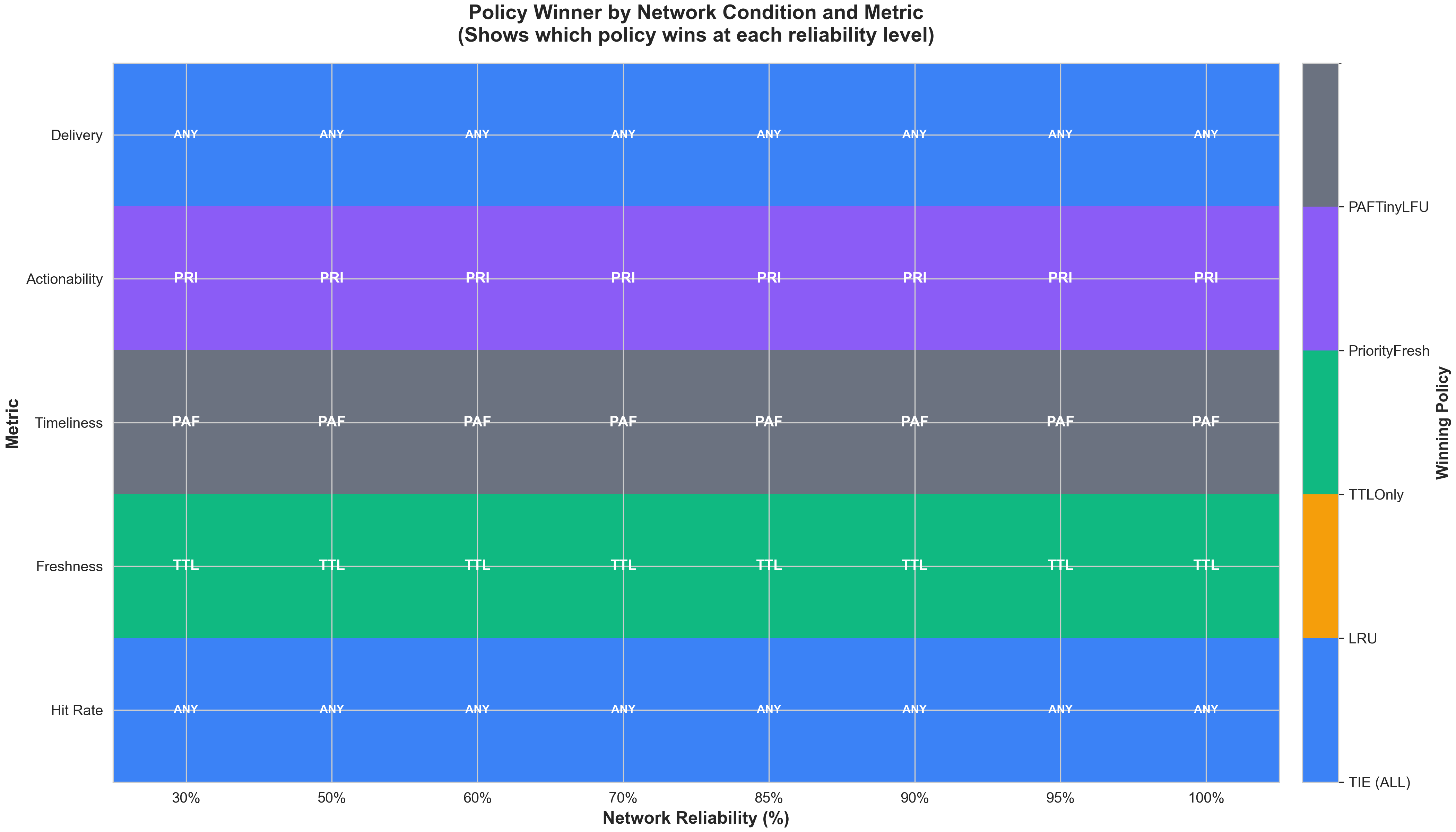}
    \caption{Network-reliability sweep: winner heatmap across metrics/policies.}
    \label{fig:network-winner-heatmap}
\end{figure}

\begin{figure}[!t]
    \centering
    \includegraphics[width=\textwidth]{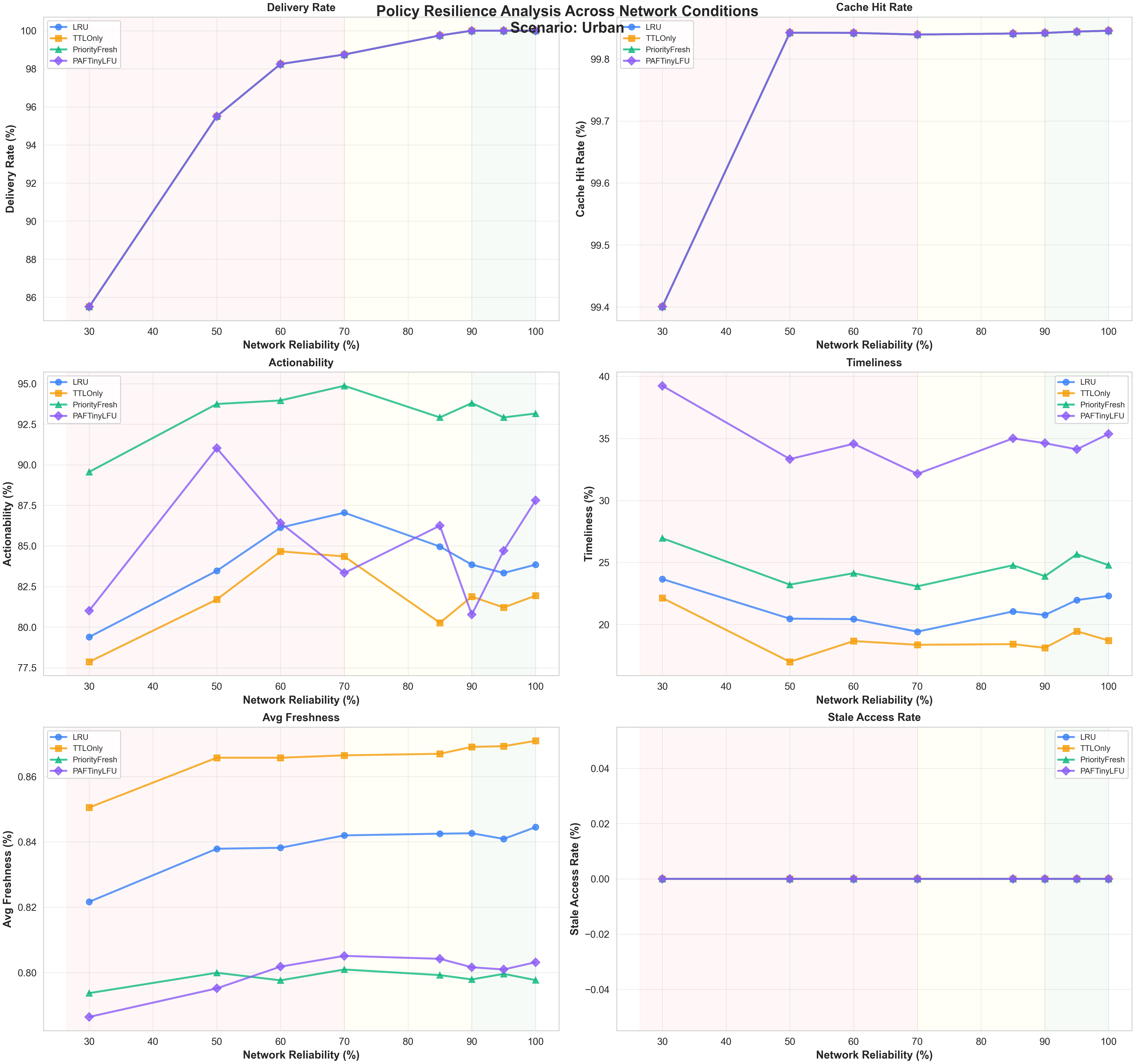}
    \caption{Network-reliability sweep: all-metrics grid overview.}
    \label{fig:network-all-grid}
\end{figure}

From \texttt{network-comparison-*.csv} and Fig.~\ref{fig:network-winner-heatmap} and Fig.~\ref{fig:network-all-grid}:
\begin{itemize}
    \item \textbf{Delivery rate tracks reliability uniformly.} With pushes disabled, all policies share essentially identical delivery under a given baseline reliability.
    \item \textbf{PriorityFresh sustains actionability-first across regimes.} The advantage in actionability-first holds from Perfect to Disaster networks.
    \item \textbf{Timeliness consistency.} PAFTinyLFU frequently attains the highest timeliness consistency, especially in mid-range reliabilities; PriorityFresh remains competitive and above LRU/TTL.
\end{itemize}

\subsubsection*{Policy resilience across network conditions (Fig.~\ref{fig:network-all-grid})}
	\textbf{Scenario.} Urban. The all-metrics grid evaluates how LRU, TTLOnly, PriorityFresh, and PAFTinyLFU respond to changing baseline reliability (30\,\% $\to$ 100\,\%), with the winner heatmap (Fig.~\ref{fig:network-winner-heatmap}) summarizing where each policy dominates.

	\textbf{Delivery rate.} Delivery rises monotonically with reliability for all policies, from roughly $\sim$86\% at 30\,\% to $\approx 99.8\%$ by $\gtrsim$80\,\%. No meaningful divergence appears among policies, confirming that packet loss and link-layer stability dominate outcomes; beyond $\sim$70\,\%, performance saturates, leaving little headroom for algorithmic gains.

	\textbf{Cache hit rate.} Hit rate mirrors delivery: a sharp climb between 30--50\,\% reliability, then a plateau at $\sim$99.8\%. The uniformity across policies implies cache efficiency is unaffected by reliability shifts, as all retain similar object pools despite connectivity fluctuations.

	\textbf{Actionability.} Divergence is clearest here. PriorityFresh maintains the highest and most stable actionability (\,$\sim$93--96\%\,) across reliabilities, outperforming LRU and PAFTinyLFU (\,$\sim$82--88\%\,) with TTLOnly lowest overall (\,$\sim$78--84\%\,). PriorityFresh's stability reflects context-sensitive retention: actionable, high-urgency alerts persist even when updates fail, whereas others degrade more under intermittent links.

	\textbf{Timeliness.} PAFTinyLFU leads throughout (\,$\sim$35--40\%\,), providing the strongest temporal stability. PriorityFresh holds moderate timeliness (\,$\sim$22--27\%\,), balancing latency and prioritization. TTLOnly trails (\,$\sim$18--20\%\,) given its fixed refresh cadence, which penalizes responsiveness under loss. The persistent gaps indicate timeliness is governed more by caching logic than by connection stability.

	\textbf{Average freshness.} TTLOnly exhibits the steepest increase, reaching $\sim$0.90 at perfect reliability as refreshes succeed more consistently. LRU follows at $\sim$0.85; PriorityFresh and PAFTinyLFU remain lower (\,$\sim$0.80--0.83\,), reflecting the prioritization vs. freshness tradeoff. Under low reliability, TTLOnly's advantage narrows due to missed refreshes, then reasserts as the network stabilizes.

	\textbf{Stale access rate.} All policies sustain 0\% stale across reliabilities, confirming robust invalidation and verification logic even under severe packet loss.

	\textbf{Interpretation.} As reliability improves, delivery, hit rate, and freshness saturate and policies converge; network quality becomes the governing factor. At lower reliabilities, context-aware behavior matters most: PriorityFresh remains most resilient on actionability, PAFTinyLFU provides the strongest temporal stability, TTLOnly optimizes freshness, and LRU serves as a steady baseline. The separation at the low end and convergence at the high end align with the \emph{ANY} patterns in the winner summary.
\FloatBarrier

\subsection{Joint sweep (cache size \& network)}
\begin{figure}[!t]
    \centering
    \includegraphics[width=\linewidth]{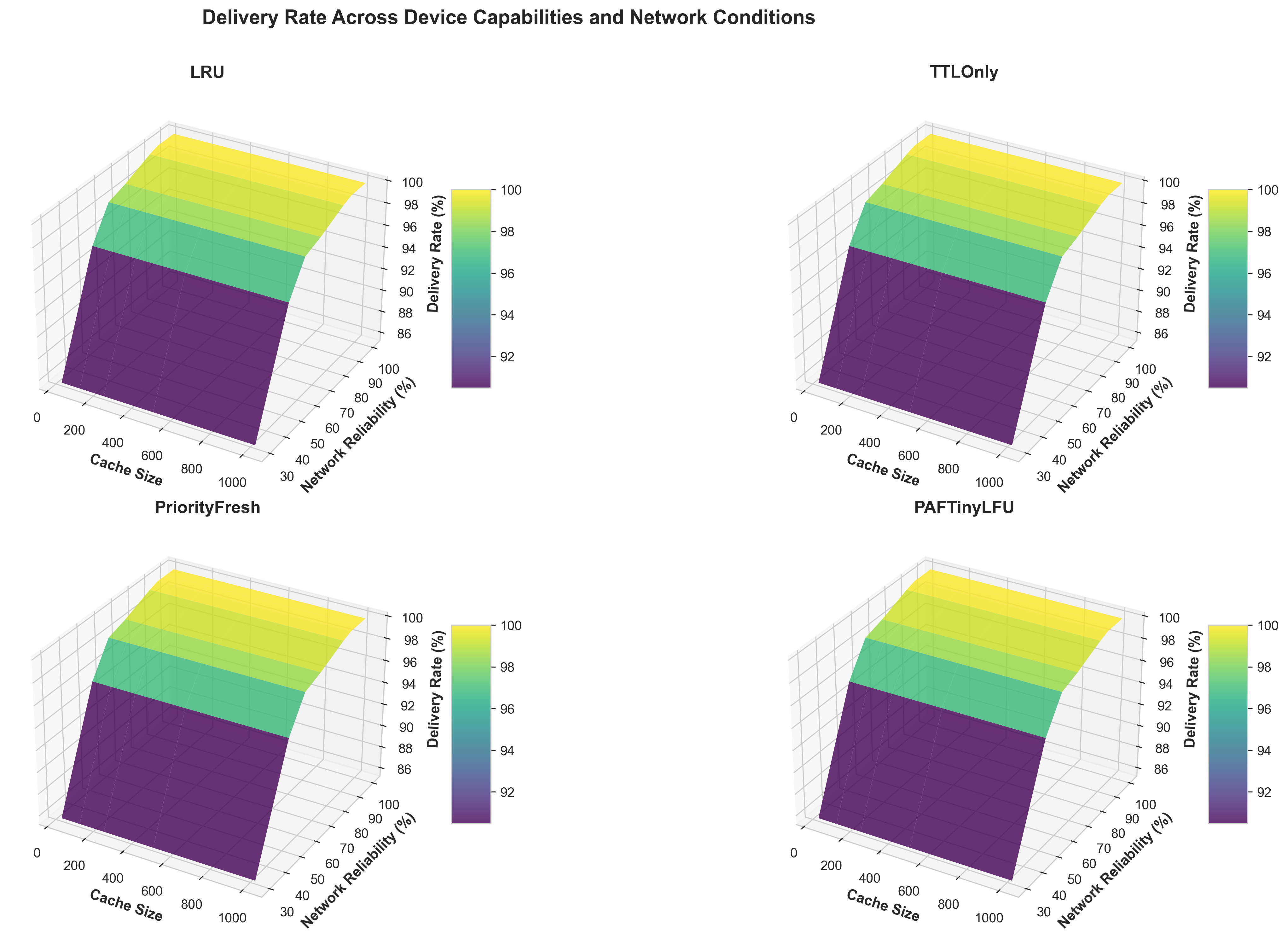}
    \caption{Joint sweep: delivery rate surface over cache size and network reliability.}
    \label{fig:combined-surface-delivery}
\end{figure}

\begin{figure}[!t]
    \centering
    \includegraphics[width=\linewidth]{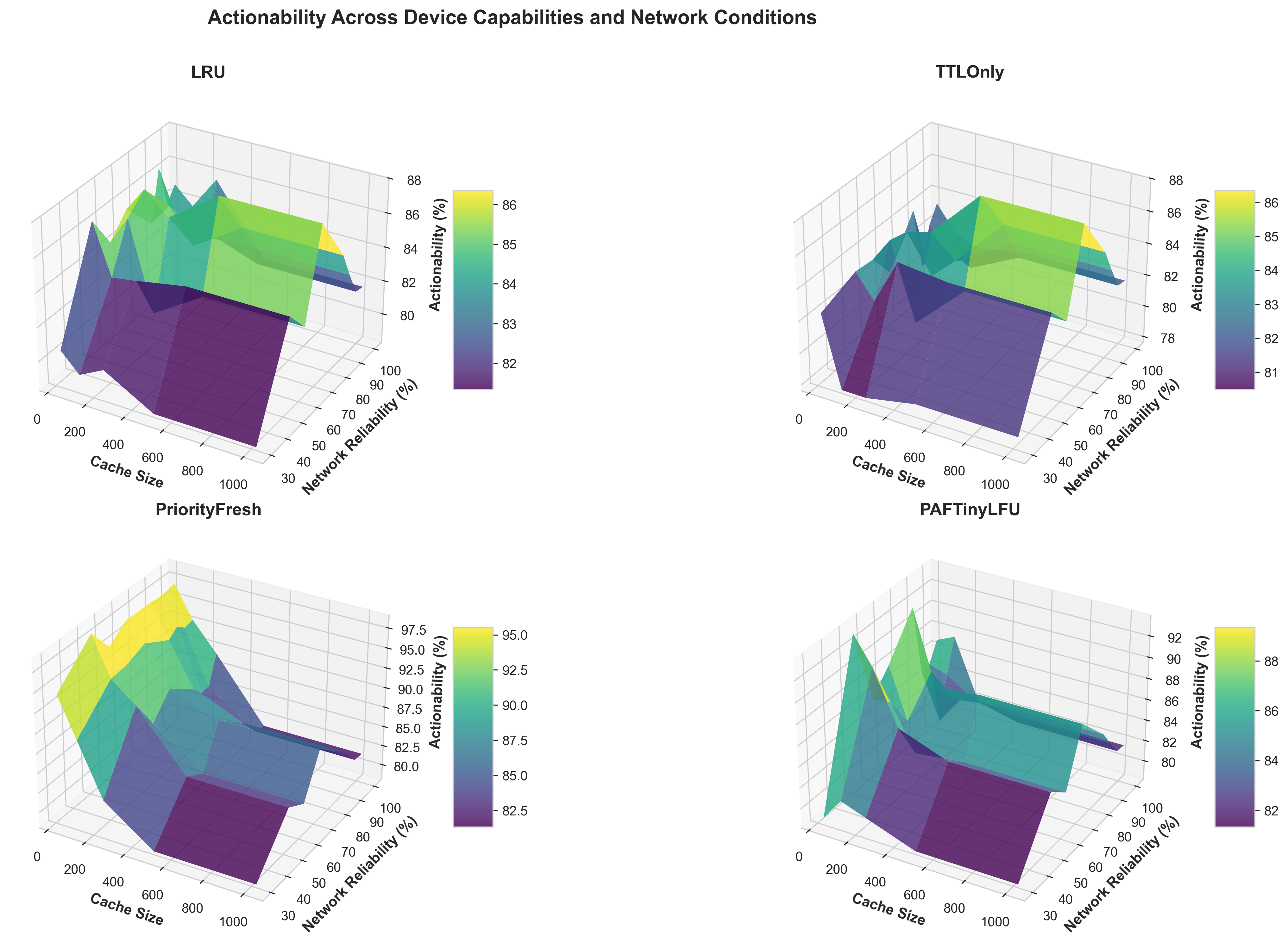}
    \caption{Joint sweep: actionability-first surface over cache size and network reliability.}
    \label{fig:combined-surface-actionability}
\end{figure}

\begin{figure}[!t]
    \centering
    \includegraphics[width=\linewidth]{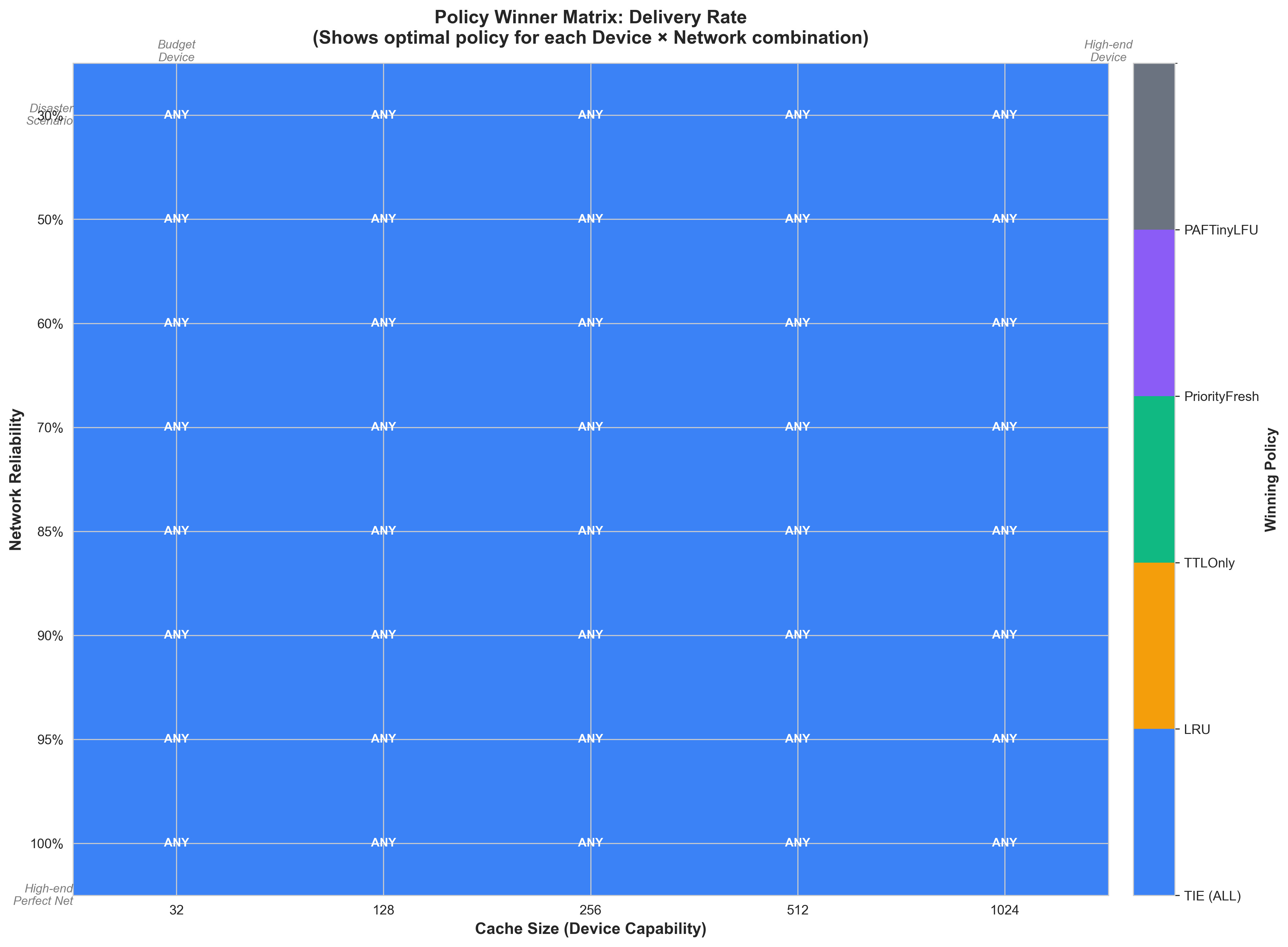}
    \caption{Joint sweep: winner cube (delivery rate).}
    \label{fig:combined-winner-cube-delivery}
\end{figure}

From \texttt{combined-comparison-*.csv} and Figs.~\ref{fig:combined-surface-delivery}, \ref{fig:combined-surface-actionability}, \ref{fig:combined-winner-cube-delivery}, and \ref{fig:combined-winner-cube-actionability}:
\begin{itemize}
    \item \textbf{Monotonic surfaces.} Delivery rises with network reliability and shows diminishing returns with larger caches.
    \item \textbf{Where policies differ.} PriorityFresh wins actionability-first across most of the grid except in high-capacity regions where all policies tie; TTLOnly leads freshness; PAFTinyLFU often wins timeliness consistency.
\end{itemize}

\subsubsection*{Actionability across device capabilities and network conditions (Fig.~\ref{fig:combined-surface-actionability})}
	\textbf{Scenario.} Urban. This 3D surface compares how actionability (rate of contextually relevant, human-usable alerts) varies with cache size (device capability) and network reliability for LRU, TTLOnly, PriorityFresh, and PAFTinyLFU.

	\textbf{LRU.} The LRU surface shows modest variation, largely within \,$\sim$82--86\%\,. Actionability rises gradually as reliability improves and plateaus once reliability exceeds $\sim$80\,\%. Cache size has minimal influence beyond $\sim$256 entries, indicating recency alone provides limited contextual prioritization; gains mainly reflect link stability rather than policy intelligence.

	\textbf{TTLOnly.} Similar range (\,$\sim$81--86\%\,). Lower values at poor reliability reflect stale drops from missed refreshes (TTL expirations require consistent updates). Even at high reliability and large caches, actionability caps below \,$\sim$87\%\,, underscoring that blind time-based eviction cannot optimize content relevance. The surface is relatively flat with a shallow slope, confirming dependence on network quality over local decision heuristics.

	\textbf{PriorityFresh.} Distinctly elevated topography, peaking near \,$\sim$97.5\%\, under high reliability and low cache. The surface gently slopes downward with increasing cache size, showing that selective triaging is most effective under constraint when the policy must choose what to retain. Performance remains relatively stable across reliability levels—often $>$\,90\% even when reliability dips below $\sim$60\,\%—highlighting context-driven optimization aligned with AWARE's focus on prioritizing actionable information over raw throughput.

	\textbf{PAFTinyLFU.} Intermediate between PriorityFresh and LRU, with peaks around \,$\sim$91--92\%\, but greater sensitivity to both cache and reliability. It performs well with small caches and good connectivity (frequency heuristics matching access patterns), but the advantage fades at large caches and unreliable networks, producing a rougher surface and indicating less contextual adaptability under constraint.

	\textbf{Interpretation.} The surfaces visualize where contextual intelligence matters most: under low cache and poor-to-moderate reliability, PriorityFresh sharply outperforms others by retaining critical alerts; as cache size and reliability increase, all policies flatten toward similar bounds (\,$\sim$85--88\%\,), demonstrating context convergence—the same \emph{ANY} phenomenon observed in winner matrices and scaling analyses. PriorityFresh's surface shows not just higher peaks but greater stability under adversity, capturing the core AWARE principle: policy differentiation is meaningful when environmental and resource constraints exist.

\subsubsection*{Delivery rate across device capabilities and network conditions (Figs.~\ref{fig:combined-surface-delivery}, \ref{fig:combined-winner-cube-delivery})}
	\textbf{Scenario.} Urban. The top surface (Fig.~\ref{fig:combined-surface-delivery}) and the bottom winner matrix (Fig.~\ref{fig:combined-winner-cube-delivery}) jointly illustrate that delivery rate remains effectively invariant across caching policies and environmental configurations, emphasizing the dominance of physical/link-layer constraints over cache decision logic.

	\textbf{3D surface analysis (top figure).} Across LRU, TTLOnly, PriorityFresh, and PAFTinyLFU, the delivery surfaces exhibit near-identical topography: (i) a sharp rise from $\sim$86\% at low reliability (\,$\approx$40\%\,) to 99--100\% once reliability exceeds 70--80\%; (ii) negligible influence from cache size, with all planes flattening into a uniform plateau beyond $\sim$256 entries; and (iii) no distinctive curvature, gradient, or deviation among policies. This confirms that delivery success is dictated by network reliability, not cache intelligence; once conditions are stable, all policies saturate at near-perfect delivery regardless of internal heuristics.

	\textbf{Winner matrix (bottom figure).} The complementary policy matrix visually reinforces this outcome: every cell of the device~$\times$~network grid is labeled \emph{ANY}, indicating no policy exhibits a measurable edge in delivery performance under any configuration. Effectively, the legend reduces to a tie \emph{(ALL)} outcome, reflecting complete metric convergence.

	\textbf{Interpretation.} Delivery rate is a non-differentiating metric across caching approaches. Caching design cannot materially affect transmission success when (1) network reliability is the limiting factor at low reliabilities, and (2) full connectivity saturates the metric at high reliabilities. While other metrics (e.g., actionability, timeliness) reveal contextual gaps between policies, delivery rate serves as a baseline indicator of infrastructure health rather than algorithmic superiority. This invariance underscores the need for context-aware evaluation—caching innovation matters when environmental uncertainty or device scarcity make delivery non-trivial.

\begin{figure}[!t]
    \centering
    \includegraphics[width=\linewidth]{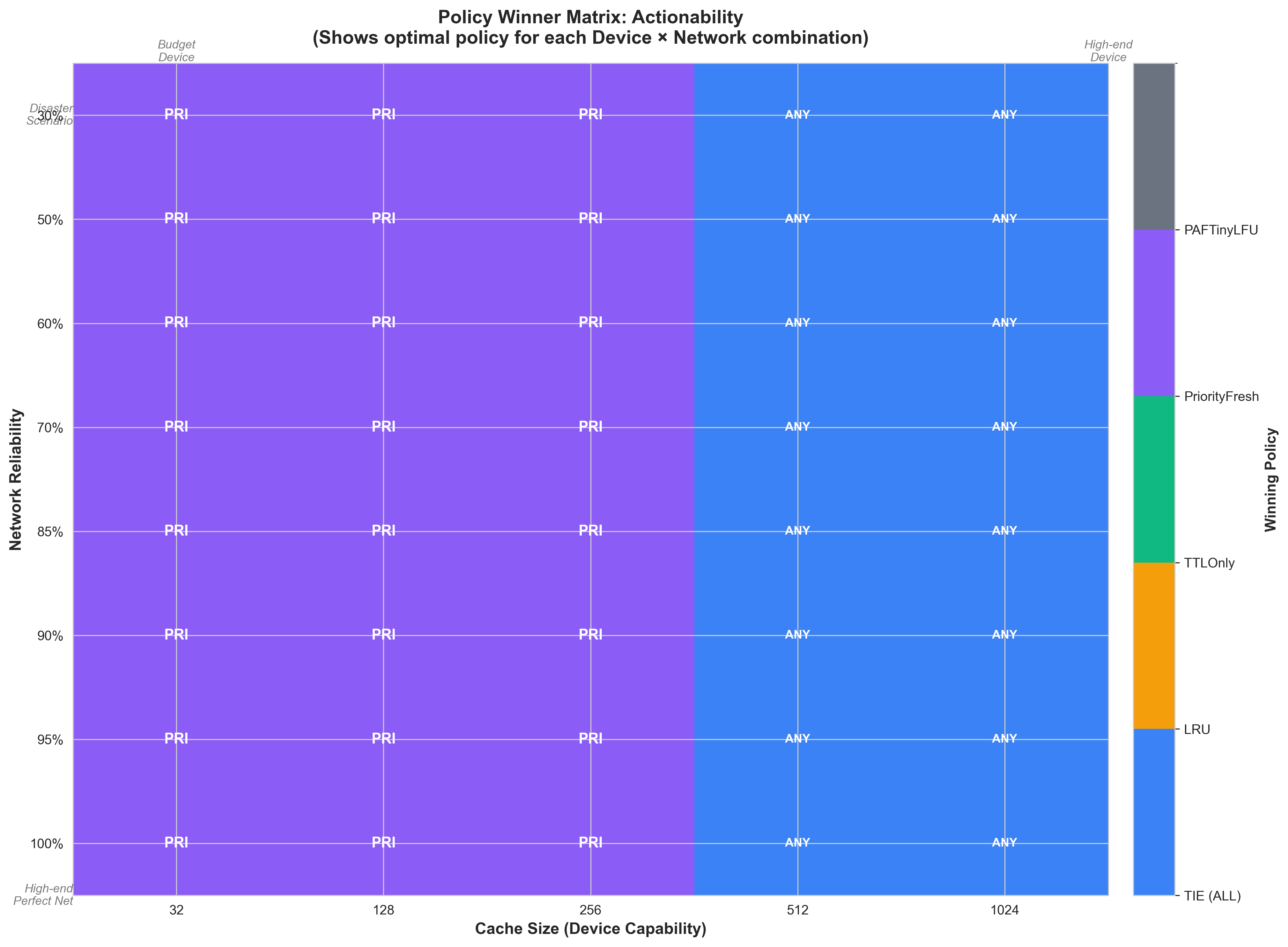}
    \caption{Joint sweep: winner cube (actionability-first).}
    \label{fig:combined-winner-cube-actionability}
\end{figure}

\subsubsection*{Policy Winner Matrix: Actionability (Fig.~\ref{fig:combined-winner-cube-actionability})}
	\textbf{Axes.} X-axis: cache size (device capability), spanning 32\,$\to$\,1024 entries. Y-axis: baseline network reliability from 0.50 to 1.00. Colors: Purple (\emph{PRI}) denotes PriorityFresh as winner; Blue (\emph{ANY}) denotes no clear dominant policy (policies perform equivalently). Right-legend bars summarize how often each winner class appears across the matrix.

	\textbf{Observed trends.} Across reliabilities and at lower-to-mid capacities (32--256), PriorityFresh (\emph{PRI}) dominates, consistently yielding the highest actionability-first performance. As capacity grows beyond $\sim$512 entries, the matrix transitions to a uniform \emph{ANY} region, indicating that no single policy retains a measurable advantage when cache and link conditions exceed threshold levels.

	\textbf{Interpretation.} This indicates a saturation point in the system's policy sensitivity: (1) At small/moderate caches, eviction/prioritization logic materially affects which alerts remain available, giving PriorityFresh an edge. (2) Beyond $\sim$512 entries, caches are large enough to retain nearly the full working set, so all policies converge—every alert remains locally accessible and differences in prioritization rarely surface as metric gains. The \emph{ANY} region is thus a performance plateau with diminishing marginal utility from smarter caching once hardware and links are abundant.

	\textbf{Relation to the research question.} The matrix evaluates raw actionability independent of operational context; under ideal, laboratory-like conditions, sophisticated prioritization yields little advantage (\emph{ANY}). In realistic emergencies—where bandwidth, reliability, and cache resources fluctuate—PriorityFresh's semantics regain importance. This underscores the need for context-aware evaluation: policy choice matters most in constrained, dynamic settings, precisely the regimes AWARE targets.




\begin{table}[t]
\centering
\caption{Context-aware recommendations by device~$\times$~network and objective (qualitative)}
\label{tab:context-reco}
\begin{tabularx}{\textwidth}{@{}l X l@{}}
        \toprule
Decision node & Branch conditions & Recommended policy \\
\midrule
\multicolumn{3}{l}{\textbf{Objective-driven overrides}} \\
\addlinespace[0.25em]
Objective: Strict Freshness SLA & Hard bounds on staleness/compliance; simple time-based behavior preferred & \textbf{TTLOnly} \\
Objective: First-Surface Regularity & Mission-critical stability of first surfaced item timing across threads & \textbf{PAFTinyLFU} \\
\addlinespace[0.5em]
\multicolumn{3}{l}{\textbf{Device $\times$ network branches (default objective: actionability-first)}} \\
\addlinespace[0.25em]
Low cache + Poor net & Constrained device, unstable link & \textbf{PriorityFresh (\emph{PRI})} \\
Low cache + Good net & Constrained device, stable link & \textbf{PriorityFresh (\emph{PRI})} \\
High cache + Poor net & Ample device, unstable link & \textbf{PriorityFresh (\emph{PRI})} \\
High cache + Good net & Ample device, stable link & \textbf{PriorityFresh (\emph{PRI})} \\
\bottomrule
\end{tabularx}
\end{table}

	\textbf{Interpretation.} Unlike raw performance matrices that converge to \emph{ANY} at large caches and high reliabilities, this decision chart integrates the contextual priorities central to AWARE. From this lens: (i) PriorityFresh is preferred across scenarios because it balances freshness, timeliness, and relevance to support human decision readiness, not just throughput; (ii) even where raw metrics equalize, contextual weighting favors PriorityFresh for consistently prioritizing critical alerts; (iii) under resource scarcity or disaster conditions, PriorityFresh's actionability-first bias yields maximal benefit by ensuring essential alerts persist when other policies focus on generic efficiency metrics.

        \textbf{Summary.} The table is a context-aware guide, not a raw scoreboard. It highlights that raw efficiency alone fails to capture operational value once infrastructure is sufficient. Under AWARE's actionability-first framing, PriorityFresh is preferred across device and network conditions; when operator objectives shift, the first two rows capture contexts in which PAFTinyLFU (first-surface regularity) or TTLOnly (strict freshness SLA) are recommended. As capacity and reliability increase, \emph{ANY} becomes acceptable due to metric convergence.
\FloatBarrier

\subsection{Extreme scenario comparison}
\label{sec:extreme-scenarios}
\noindent\textbf{Context.} All four cache policies (LRU, TTLOnly, PriorityFresh, PAFTinyLFU) are evaluated under four edge-case combinations of device capability and network quality:
\begin{enumerate}
    \item \textbf{Best case:} High-end device with a near-perfect network.
    \item \textbf{Good device / poor network.}
    \item \textbf{Budget device / good network.}
    \item \textbf{Worst case:} Budget device with disaster-grade network.
\end{enumerate}

\begin{figure}[!t]
    \centering
    \includegraphics[width=\textwidth]{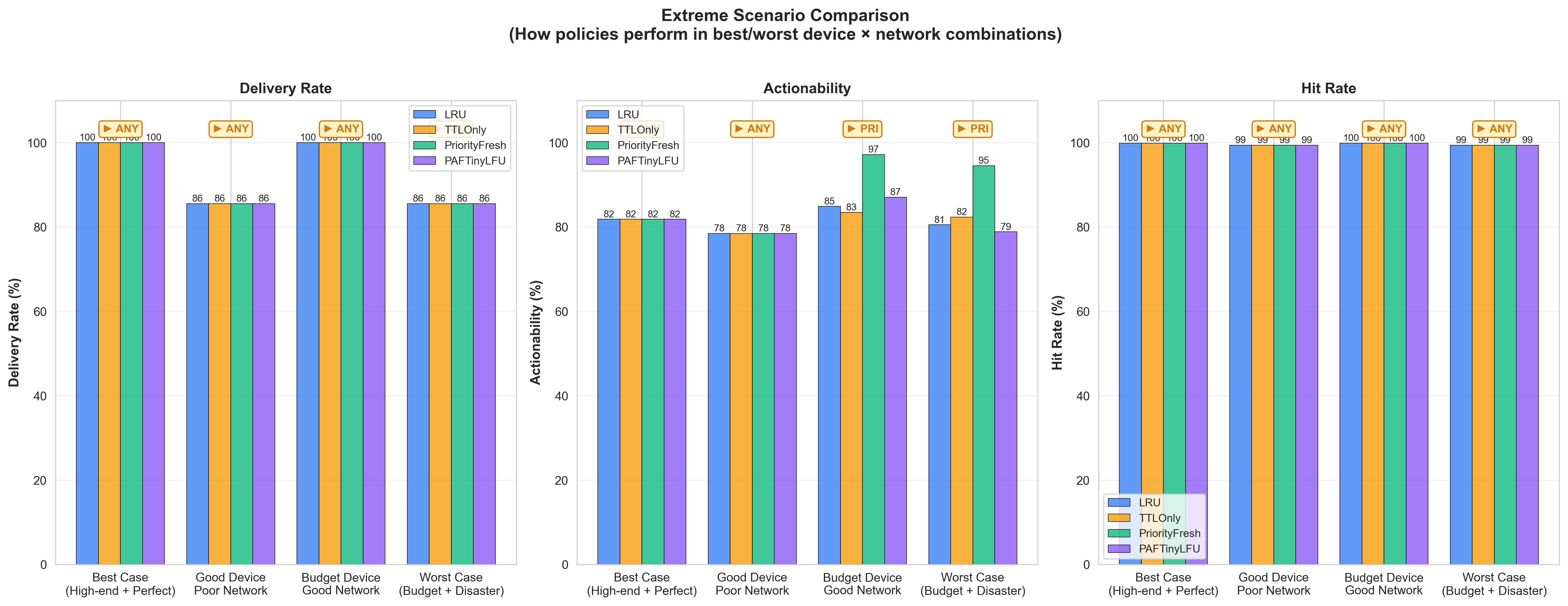}
    \caption{Extreme scenario comparison: four device/network corners summarizing delivery, hit rate, and actionability across policies. Bars labeled \emph{ANY} indicate convergence where no policy holds a measurable advantage.}
    \label{fig:combined-extreme-scenarios}
\end{figure}

\subsubsection*{Delivery rate}
Across all scenarios, delivery is invariant across policies: \textbf{100\%} in the best case and other high-reliability settings, and \textbf{$\approx 86\%$} under either network degradation or low-end hardware (scenarios 2--4). Each policy experiences identical reductions, implying that delivery is constrained by physical network reliability rather than algorithmic caching. This uniformity aligns with the \emph{ANY} regions summarized by the winner matrices (e.g., Fig.~\ref{fig:combined-winner-cube-actionability}), where no strategy exhibits a measurable advantage once transmission success is the bottleneck.

\subsubsection*{Actionability}
This dimension differentiates policies most clearly under constraint. PriorityFresh (\emph{PRI}) leads in the two constrained scenarios:
\begin{itemize}
    \item \textbf{97\%} actionability for Budget + Good Network.
    \item \textbf{95\%} for Worst Case (Budget + Disaster).
\end{itemize}
Other methods cluster between \,$\sim$79--87\%\,. The pattern highlights that PriorityFresh dynamically prioritizes context-relevant items even when throughput and device capability are limited. By contrast, the nearly homogeneous \,$\sim$82\%\, actionability under the Best Case shows that, when conditions are optimal, all policies converge—again reinforcing that scarcity is what triggers meaningful divergence.

\subsubsection*{Hit rate}
Hit rate mirrors delivery at \,$\sim$99--100\%\, across all conditions and policies. The near-perfect values indicate that even the smallest caches considered contain the working set for the simulated data volume. Consequently, eviction details have negligible effect under both ideal and degraded networks, yielding another \emph{ANY} zone.

\subsubsection*{Interpretation}
Algorithmic advantages emerge when resource scarcity exists in the dimension a policy targets. In resource-abundant contexts (ample cache, stable network), policies behave equivalently because the system is already saturated with available data. In resource-constrained contexts (budget hardware or unstable connectivity), PriorityFresh retains a distinct edge by selectively preserving high-impact alerts—an effect not visible in raw hit or delivery, but captured by actionability. Thus, the \emph{ANY} labels reflect convergence under ideal or uniformly constrained regimes where performance is dominated by network physics rather than cache intelligence. Differentiation appears when simulations incorporate contextual stressors, directly tying back to the research question on context-aware smart caching versus purely performance-oriented caching.

\subsection{Timeline behavior}
\begin{figure}[!t]
    \centering
    \includegraphics[width=\textwidth]{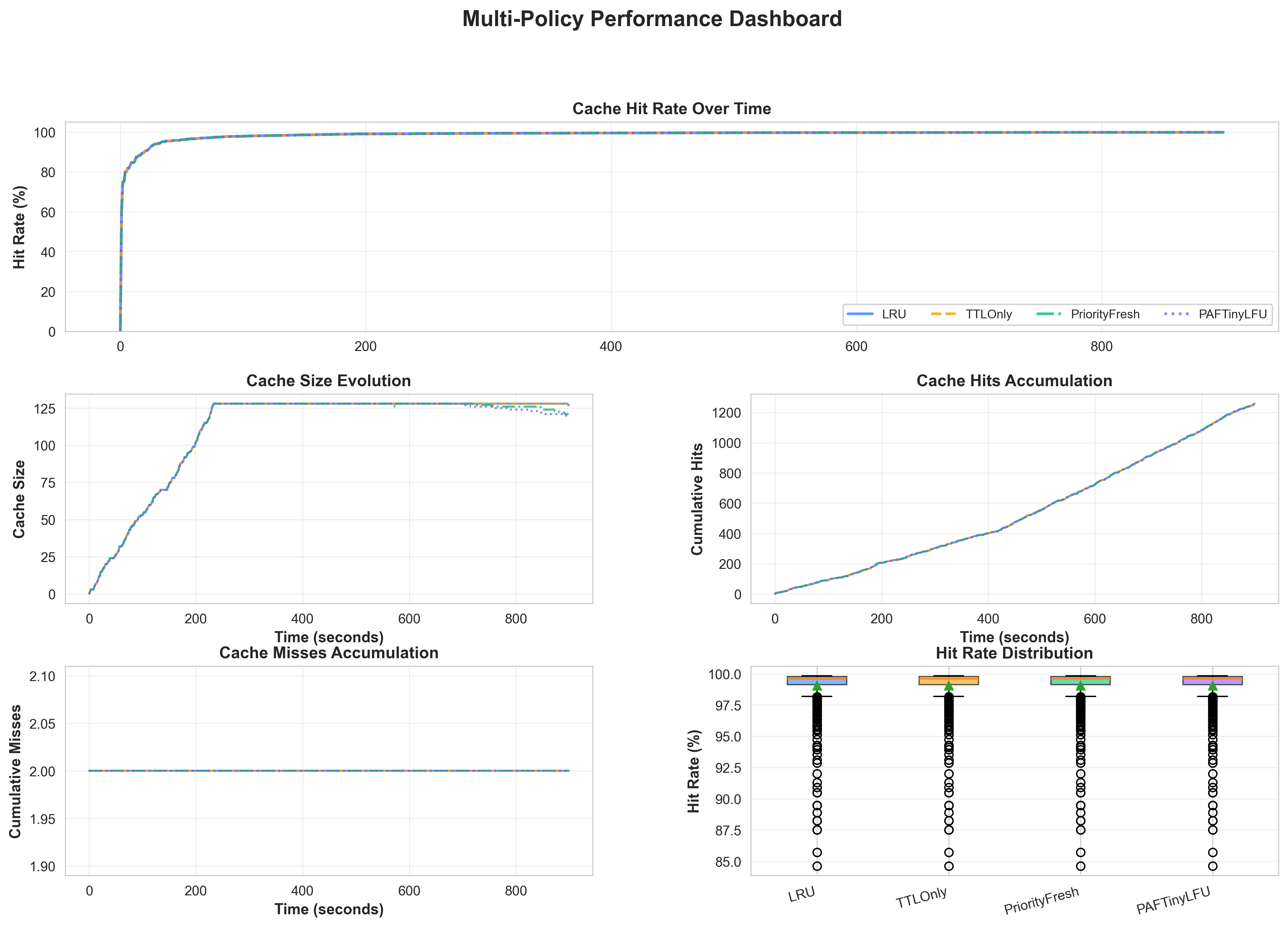}
    \caption{Timeline dashboard view (cache size, hits/misses, and derived hit rate over run time).}
    \label{fig:timeline-dashboard}
\end{figure}

The per-run timeline (Fig.~\ref{fig:timeline-dashboard}) shows smooth convergence of hit rate to $\sim$0.998 for all policies under the baseline seed, consistent with the point-in-time summaries.

\subsubsection*{Multi-policy performance dashboard (Fig.~\ref{fig:timeline-dashboard})}
    	\textbf{Cache hit rate over time.} All policies converge to nearly identical curves, climbing from 0\,\% to $\approx 99.8\%$ within the first $\sim$150\,s and remaining stable thereafter. The steep rise then plateau indicates rapid steady-state cache saturation; once the working set is cached, policy logic no longer affects aggregate hit rate.

    	\textbf{Cache size evolution.} Each trace increases linearly until reaching the 128-entry capacity at roughly $t\,\approx\,200\,\mathrm{s}$, then oscillates minutely around the limit. Overlapping trajectories confirm identical admission/eviction dynamics during warm-up, reinforcing comparable fill efficiency across policies under this workload.

    	\textbf{Cache hits accumulation.} Cumulative hits grow linearly and overlap across policies, reaching $\sim 1,200$ hits by 900\,s. This uniformity signals consistent throughput and retrieval frequency, implying that network delay or device constraints, not cache logic, dictate servicing pace once stabilized.

    	\textbf{Cache misses accumulation.} All lines flatten at $\approx 2$ total misses after warm-up, indicating statistically negligible misses thereafter. Near-zero miss accumulation highlights that the cache fully captures the active dataset, validating that 128 entries suffice for the tested demand profile.

    	\textbf{Hit-rate distribution.} Boxplots cluster tightly around 99--100\%, with median/mean nearly identical and only a few low outliers near 85--90\%. The narrow spread and symmetry confirm temporal stability—no policy shows degradation or transient volatility once equilibrium is reached.

    	\textbf{Interpretation.} Across temporal and cumulative views, the dashboard illustrates steady-state convergence: after brief initialization, all four algorithms perform equivalently. Under stable urban conditions and modest cache size, locality and access pattern dominate, leaving minimal room for algorithmic differentiation. This supports that policy advantages emerge primarily under dynamic or resource-constrained contexts, not during steady, fully saturated operation.

\FloatBarrier

\section{Discussion and conclusion}
\subsection{Takeaways}
PriorityFresh preserves system efficiency while consistently improving the actionability of what users see first, particularly when cache is constrained or networks are degraded. TTLOnly is a strong baseline for freshness; PAFTinyLFU provides the most stable first-push timing. In practice, operators can tune PriorityFresh via weights $(w_S,w_U,w_F)$ and, if pushes are enabled, a simple base-score threshold $\theta$ and dedup/rate limits—no ML coupling is required.

\subsection{Answering the Research Question}
\label{sec:rq-answer}
Findings suggest that the most effective way to deliver the most crucial alerts is to optimize what surfaces first rather than maximizing total coverage. PriorityFresh is designed for this exact objective:
\begin{itemize}
    \item \textbf{Actionability-first by construction.} The scoring emphasizes urgency, severity, and recency using fixed, transparent weights. This yields the top actionability-first ratio across regimes without ML inputs.
    \item \textbf{Hold until no longer relevant.} Decay and expiry remove items once they are no longer pressing. This maintains focus on ongoing, high-impact threads instead of churning through the catalog.
    \item \textbf{Tradeoffs are intentional.} PriorityFresh is not a mass-coverage model. Its average freshness and timeliness stability reflect a conscious trade to ensure the right items show up first. TTLOnly leads in average freshness; PAFTinyLFU stabilizes timing—both are compatible fallbacks when those criteria are primary.
    \item \textbf{Operator levers.} Weights $(w_S,w_U,w_F)$ and, if pushes are enabled, the threshold $\theta$, act as knobs: increase $w_U$/$w_S$ or $\theta$ when “crucial alerts” should be filtered aggressively; relax them when broader recall or freshness is preferred.
    \item \textbf{Push discipline (when enabled).} Rate limits and deduplication suppress redundant pings. The same actionability-first principle can govern which threads break through to notifications using the base score.
\end{itemize}

If the research question is about surfacing what matters most, the evaluation indicates PriorityFresh answers it: it consistently leads on actionability-first without sacrificing system efficiency, and it provides explicit controls to tune how aggressively “pressing” is interpreted for different operational contexts.


\subsection{Conclusion}
 AWARE reframes client-side alert delivery around \emph{what should surface first}, not just how many alerts are delivered. Within a reproducible, simulation-only environment across cache sizes, network reliabilities, and joint extremes, the following patterns hold: (i) delivery and hit rate saturate and are largely governed by network conditions, (ii) policy differences emerge under constraint, where PriorityFresh improves actionability without sacrificing efficiency, and (iii) PAFTinyLFU provides the most stable timing, while TTLOnly maximizes freshness. The PF model, when present, serves only to generate reasonable alert streams and remains decoupled from PriorityFresh.

\noindent\textbf{Implications.} For operators, the recommendation table and winner views offer practical guidance: under scarce resources or degraded links, prioritize semantics (PriorityFresh); when timing regularity is paramount, consider PAFTinyLFU; when average freshness is the sole objective, TTLOnly remains a simple baseline. As capacity and reliability increase, policies converge (\emph{ANY}), signaling diminishing returns from more sophisticated caching.

\subsection*{Limitations and future work} This evaluation is simulation-only and assumes idealized alert distributions. Future work will integrate live network traces, user-interaction modeling, and ethical safeguards against alert suppression bias. The following open items are known limitations, nuances, and margins for error to be addressed in subsequent experiments:
\begin{itemize}
    \item \textbf{Multi-seed variance and confidence intervals.} Report multi-seed runs with confidence intervals to characterize variance/robustness beyond single-seed behavior.
    \item \textbf{Sensitivity of weights and decay.} Systematically sweep $(w_S, w_U, w_F)$ and the decay rate $\lambda$; identify stability regions versus brittle regimes.
    \item \textbf{Push-channel behavior.} When $R$, $D$, and $\theta$ are enabled, study duplicate storms, rate-limit interactions, and non-suppressible Immediate/Extreme cases.
    \item \textbf{Update-storm stress tests.} Evaluate rapid correction chains and contradictory thread sequences to assess ordering, dedup, and surfacing stability under bursty updates.
    \item \textbf{CAP mis-coding impacts.} Quantify how mis-labeled severity/urgency in CAP affects PriorityFresh ordering and actionability metrics.
    \item \textbf{Geofencing error tolerance.} Measure sensitivity to boundary jitter, degraded GPS, and OS-dependent geofencing precision.
    \item \textbf{PF generator validation.} Validate the PF generator against real CAP distributions (inter-arrivals, escalation rates, thread lengths) for external realism.
    \item \textbf{Real-environment \& historical validation.} Replay historical CAP archives and real alert corpora; integrate live network telemetry (carrier/APNs/FCM delivery timing, geofence precision) and field pilot data to ground simulation parameters and detect divergence from in-situ behavior.
    \item \textbf{User-interaction model.} Incorporate delayed reads, dismissals, and cognitive-load effects to understand human-in-the-loop dynamics.
    \item \textbf{Device/energy profiling.} Profile CPU/memory/storage footprint and power on budget devices to bound overheads and tune defaults.
    \item \textbf{DTN/mesh scenarios.} Extend to intermittent, multi-hop connectivity models.
    \item \textbf{Cross-hazard/locale generalization.} Test floods, wildfires, severe storms, and multilingual templates to assess generality.
    \item \textbf{Ethical safeguards and fail-open.} Specify and test fail-open guarantees for high-impact alerts (Immediate/Extreme).
    \item \textbf{Formal analysis.} Analyze eviction ordering and exponential-decay behavior formally to bound worst-case outcomes.
\end{itemize}

\section{Acknowledgements}
This work was carried out internally within the Floodwatch project at the University of Virginia. Special thanks to Floodwatch undergraduate team members for practical discussions on alert operations, simulator design and presentation, and general research direction.

\FloatBarrier

\end{document}